
\input phyzzx
\hfuzz 50pt
\font\mybb=msbm10 at 10pt

\def\Bbb#1{\hbox{\mybb#1}}

\def\bT{\Bbb {T}}
\def\bR{\Bbb {R}}

\def\bfomega{\omega\kern-7.0pt
\omega}

\magnification=950
\REF\gary{G.W.
Gibbons, G. Papadopoulos 
and K. Stelle, {\sl HKT and OKT
Geometries on Soliton
Black Hole 
Moduli Spaces}, Nucl. Phys. {\bf B508} (1997)
623,
hep-th/9706207.}
\REF\micha{J. Michelson and 
A. Strominger, {\sl
Superconformal 
Multi-Black Hole
Quantum Mechanics}, HUTP-99/A047,
hep-th/9908044.\break
R. Britto-Pacumio, J. Michelson, 
A. Strominger and A.
Volovich,
{\sl Lectures on superconformal 
quantum mechanics and multi-black
hole 
moduli spaces}, hep-th/9911066.}
\REF\gutpap{J. Gutowski and 
G.
Papadopoulos, {\sl The dynamics of very special black holes}, 
Phys. Lett. {\bf
B472} (2000) 45; hep-th/9910022 .}
\REF\mss{A. Maloney, M. Spradlin 
and A.
Strominger, {\sl Superconformal
Multi-Black Hole Moduli Spaces 
in Four
Dimensions}, hep-th/9911001.}
\REF\gutpapa{J. Gutowski and G. Papadopoulos,
{\sl Moduli spaces for four-dimensional and five-dimensional black holes},
hep-th/0002242.}
\REF\sumati {D. Marolf and S. Surya, {\sl On the Moduli space
of the localised 1-5 system}, hep-th/0005146.}
\REF\deg{Y. Degura and  K. Shiraishi, {\sl
 Effective field theory of slowly moving 'extreme black holes'},
 hep-th/0006015.} 
\REF\gibr {G.W.Gibbons and P.J.Ruback,{\sl
The motion of extreme Reissner-Nordstr\"om black
holes in the low velocity limit}, Phys. Rev.
Lett. {\bf 57} (1986) 1492.}
\REF\fere {R.C.Ferrell \& D.M.Eardley,{\sl
Slow motion scattering and coalescence of
maximally charged black holes}, Phys. Rev. Lett. {\bf 59} (1987)
1617.}
\REF\shiraishi{K. Shiraishi, {\sl Moduli space metric for
maximally-charged
dilaton black holes}, Nucl. Phys. {\bf B402} (1993)
399.}
\REF\dps{M. Douglas, J. Polchinski and A. Strominger, {\sl Probing
five-dimensional black holes with D-branes}, JHEP {\bf 9712} (1997) 003,
hep-th/9703031.}
\Ref\mald{J. Maldacena, {\sl Branes probing black holes}, 
Nucl.Phys.Proc.Suppl. {\bf 68} (1998) 17, hep-th/9709099.}
\REF\ahor{O.
Aharony, J. Sonnenschein, S. Yankielowicz and S. Theisen, {\sl Field theory
questions for string theory answers}, Nucl.Phys. {\bf B493} (1997) 177,
hep-th/9611222.}
\REF\dls{M. Douglas, D.A. Lowe and J.H. Schwarz, {\sl Probing
F-theory 
with multiple branes}, Phys.Lett. {\bf B394} (1997) 297,
hep-th/9612062.}
\REF\ss{ J. Scherk and J. H. Schwarz {\sl How to get masses 
from extra dimensions},  Nucl.Phys. {\bf B153} (1979) 61.}
\REF\BT{E. Bergshoeff and P.K. Townsend, {\sl  Solitons on the
supermembrane},  JHEP {\bf 9905} 021 (1999), hep-th/9904020.} 
\REF\bergpapa{ E. Bergshoeff, M. de Roo, M.B. Green,
 G. Papadopoulos, P.K. Townsend
{\sl Duality of type II 7 branes and 8 branes},
 Nucl.Phys. {\bf B470} (1996) 113; hep-th/9601150}
\REF\pope{ I.V. Lavrinenko, H. Lu and C.N. Pope 
{\sl Fiber bundles and generalized dimensional reduction},
 Class.Quant.Grav. {\bf 15} (1988) 2239; hep-th/9710243.}
 \REF\gukov{S. Gukov, C. Vafa and  E. Witten {\sl
  CFT'S from Calabi-Yau four-folds}, hep-th/9906070.}
\REF\lambert{J.M. Izquierdo, N.D. Lambert, G. Papadopoulos and P.K. Townsend,
{\sl Dyonic Membranes}, Nucl. Phys. {\bf 460} (1996) 560, hep-th/9508177.}
\REF\russo{J.G. Russo and  A.A. Tseytlin,
 {\sl Waves, boosted branes and bps states in M-theory}
  Nucl.Phys. {\bf B490} (1997) 121; hep-th/9611047.} 
\REF\costa{M. Costa and G. Papadopoulos, {\sl Superstring dualities
and p-brane bound states}, Nucl. Phys. {\bf B510} (1998) 217, hep-th/9612204.}
  \REF\ds {D-E Diaconescu and N. Seiberg, {\sl The Coulomb Branch
of (4,4) Supersymmetric Field Theories in Two Dimensions}, JHEP {\bf 9707}
(1997) 001, hep-th/9707158.}
\REF\is{K. Intriligator and N. Seiberg, {\sl
Mirror symmetry in three dimensional gauge theories}, Phys. Lett. {\bf B387}
(1996) 513, hep-th/9607207.} 
\REF\lwy{K. Lee, E. Weinberg and P. Yi, {\sl The
moduli space of 
many BPS monopoles for arbitrary gauge groups}, Phys.Rev. {\bf
D54} 
(1996) 1633, hep-th/9602167.}
\REF\gibb{  G.W. Gibbons and  P. Rychenkova
{\sl  Hyperk\"ahler quotient construction of BPS monopole moduli spaces},
hep-th/9608085.} 
\REF\dkmtv{N. Dorey, V.V. Khoze, M.P. Mattis, D. Tong and S.
Vandoren, 
{\sl Instantons, three-dimensional gauge theory, and the
Atiyah-Hitchin 
manifold}, Nucl.Phys. {\bf B502} (1997) 59,
hep-th/9703228.}
\REF\cd{C. Fraser and D. Tong, {\sl Instantons, three
dimensional gauge  theories and monopole moduli spaces}, Phys.Rev. {\bf D58}
(1998) 085001, 
hep-th/9710098.}
\REF\agf{L.
Alvarez-Gaum\'e and D. Freedman, {\sl Potentials for the 
supersymmetric
nonlinear sigma model}, Commun.Math.Phys.{\bf 91:87} (1983).}
\REF\gpaptown{G.
Papadopoulos and P.K. Townsend,
{\sl Massive (p,q)-supersymmetric sigma models
revisited}, Class.Quant.Grav.{\bf 11}(1994) 2163 , hep-th/9406015.}
\REF\becker{K. Becker, M. Becker and A. Strominger,
{\sl Five-branes, membranes and nonperturbative string theory},
Nucl. Phys. {\bf B456} (1995) 130, hep-th/9507178.}  
\REF\kallosh{E. Bergshoeff, R. Kallosh, T. Ort\'in and G. Papadopoulos,
{\sl Kappa symmetry, supersymmetry and intersecting branes}, 
Nucl. Phys. {\bf B502} (1997) 149.}
\REF\stromk{M. Marino, R. Minasian, G. Moore and A. Strominger,
{\sl Nonlinear Instantons from supersymmetric branes}, JHEP 0001:005 (2000),
hep-th/9912206.}
\REF\hsw{P.S. Howe, E. Sezgin and P.C. West,
{\sl Covariant Field Equations of the M-theory 5-brane},
Phys. Lett. {\bf B399} (1997) 49; hep-th/9702118.}
\REF\douglas{M. Douglas, {\sl Branes within Branes}, hep-th/9512077.}
\REF\pol{J. Polchinski, {\sl TASI Lectures on D-branes}, hep-th/9611050.}
\REF\kinky{N.
Lambert and D. Tong, {\sl Kinky D-strings} Nucl.Phys. {\bf B569} (2000) 606,
hep-th/9907098.}
\REF\keshav{S. Chakravarty, K. Dasgupta, O. Ganor and G. Rajesh, {\sl 
Pinned branes and new non lorentz invariant theories}, hep-th/0002175.}
\REF\atseytlin{A.A. Tseytlin, {\sl No force condition
and BPS combinations of p-branes in eleven- and ten-dimensions},
Nucl. Phys. {\bf B487} (1997) 141; hep-th/9609212.}
\REF\sw{N. Seiberg and E. Witten, {\sl String
theory and noncommutative 
geometry}, JHEP {\bf 9909} (1999) 032,
hep-th/9908142.}
\REF\abs{O. Aharony, M. Berkooz, N. Seiberg, {\sl Light-cone
description 
of (2,0) superconformal theories in six dimensions}, 
Adv.Theor.Math.Phys. {\bf 2} (1998) 119, hep-th/9712117.} 
\REF\sked{H. J. Boonstra, B. Peeters and  K. Skenderis  
{\sl Duality and asymptotic geometries}
 Phys.Lett. {\bf B411} (1997) 59; hep-th/9706192.} 

\Pubnum{ June 2000\vbox{\hbox{}\hbox{} }
}
\date{KCL-TH-00-34}
\pubtype{}
\titlepage
\title{Brane Potentials and Moduli Spaces}
\author{George Papadopoulos${}^{1,2}$ and
David Tong${}^{2}$}
\ 
\address{${}^1$Theory Division, CERN, 1211, Geneva 23,
Switzerland}
\address{$^{2}$Department of Mathematics, Kings College London,
\break Strand, London WC2R 2LS, UK\break
{}\break
{\tt gpapas,
tong@mth.kcl.ac.uk}}
\pagenumber=1


\def\C{\mkern1mu\raise2.2pt\hbox{$\scriptscriptstyle|$}\mkern-7mu{\rm
C}}

\def\tt{\theta}

\def\ft#1#2{{\textstyle{{\scriptstyle #1}\over
{\scriptstyle #2}}}}

\font\mybb=msbm10 at 10pt

\def\Bbb#1{\hbox{\mybb#1}}

\def\bT{\Bbb {T}}
\def\bR{\Bbb {R}}

\def\bZ {\Bbb{Z}}
\def\bfomega{\omega\kern-7.0pt
\omega}

\abstract{It is shown that the supergravity moduli spaces of D1-D5 
and D2-D6 brane systems
coincide with those of
 the Coulomb branches of the 
associated non-abelian gauge
theories. We further discuss situations in which worldvolume 
brane actions include a potential term generated 
by probing certain supergravity backgrounds.
We find that in many cases, the appearance of the potential is due
to the application of the Scherk-Schwarz mechanism.
We give some examples and discuss the existence of novel 
supersymmetric brane configurations. }
\endpage

\chapter{Introduction}

Recently, there has been much progress in understanding the moduli spaces 
of multiple
black holes  in the supergravity context [\gary-\sumati]. 
For earlier work in
this area, see [\gibr-\shiraishi]. Given the emphasis 
of string theory over the
past few years, it is natural to ask whether one 
can successfully compare such
moduli spaces with those of gauge theories. 
In the first part of this paper  we point out that, in
two simple cases, the moduli spaces do 
indeed coincide with the quantum
corrected Coulomb branch of a 
corresponding non-abelian Yang-Mills theory.

While similar in spirit to the pre-AdS/CFT probe calculations [\dps,\mald],
the scenario considered here involves an important difference: rather than
restricting attention to  test-particles (branes) 
moving in a fixed
background, the background is determined by the positions of other branes which
are themselves self-gravitating and dynamical. Therefore we are necessarily
dealing with the scattering of multiple branes. Multiple D3-branes have been
treated as probes in the past [\ahor,\dls], but only in situations in which
various simplifications ensure  that the probes do not interact with each
other, and the resulting moduli space is simply the symmetric product of the
single probe moduli space. In the present case we find that interactions
between branes do occur, but are restricted to two-body forces.
Specifically,
we show that the supergravity moduli spaces of  D2-D6 and  D1-D5 brane
configurations, each of which preserve eight supercharges, coincide with the
Coulomb branches of $d=3$, ${\cal N}=4$, and $d=2$, ${\cal N}=(4,4)$,
non-abelian Yang-Mills theory respectively.
 Of course, the supergravity
and gauge theory calculations have different ranges of validity and agreement
between them points to the existence of a non-renormalisation theorem which, in
the present case, reduces to a strong constraint on the geometry of the
manifold due to the preservation of eight supercharges and the remaining
Lorentz invariance: the moduli metric is restricted to be hyperK\"ahler (HK)
for the D2-D6-brane configuration and strong hyperK\"ahler with torsion (HKT)
for D1-D5-brane configuration. As is the norm in such calculations, the
classical supergravity result is reproduced by one-loop effects in the gauge
theory. We argue that there are no non-perturbative corrections.

In the second half of this paper, we discuss situations in which 
the low-energy dynamics of branes includes a potential term. 
Such potentials may be generated in the context of compactifications 
by using the Scherk-Schwarz (SS)  mechanism [\ss]. Since most  worldvolume
actions of various branes can be related by Kaluza-Klein type
of compactifications, the SS mechanism can be used to
generate potentials on the brane\foot{In the context of the 
supergravity approach to  branes this 
has been used in [\bergpapa] and further explored in [\pope]. The 
appearance of potentials in M-theory compactifications 
with non-trivial background form field strengths have been
investigated in [\gukov].}. In particular one begins from
the standard Dirac-Born-Infeld type of action of a (D-, M-, NS-) p-brane
and after giving an appropriate expectation value
to either a {\sl transverse scalar} or to a {\sl  Born-Infeld} (BI)
 type of field or to both,
one finds after compactification in n-directions  a (p$-$n)-brane
action which has a scalar potential term.
In many cases, the scalar potential is just a constant shift
in the reduced action but if the p-brane is placed in an appropriate
supergravity background, then a non-trivial potential can appear.
The scalar potentials that appear by placing D-branes 
in a constant
B-field or in the non-trivial compactification of the
M2-brane in [\BT] are examples of this.
There are many cases that one can consider by choosing
different supergravity backgrounds and by placing various
brane probes in them. However we shall not explore
explore all these possibilities here. Instead
we shall present some new examples including compactification 
in the presence of a constant B-field and non-trivial compactification 
of D-brane worldvolume actions 
in the presence of a ten-dimensional KK-monopole.

For the cases associated with non-trivial compactifications
in a KK-monopole background,  we shall show that there is an 
alternative bulk explanation for the presence of a potential
in the (p$-$n)-brane action. In particular we shall find that
the same potential appears on a (p$-$n)-brane probe placed
in a non-marginal BPS supergravity background. Such backgrounds
were first found in [\lambert] and further explored in [\russo, \costa].

In addition we shall investigate the presence of supersymmetric solutions 
in the various probe actions with a scalar potential and  and examine 
their properties. Some of these have the interpretation of rotating 
branes. Finally, we also discuss 
the generation of potentials  in the context of supersymmetric 
gauge theories.

In the section two we investigate the D1-D5 system, and in section 
three we examine the D2-D6
system. Finally in section four we discuss brane potentials.

\chapter{The D1-D5 brane system}

\section{Supergravity}

Our starting point is the much studied
D1-D5 system, comprised of $Q_1$ D-strings lying in the 01 directions and 
$Q_2$ D5-branes lying in the 012345 directions. The metric of the associated 
supergravity solution is, 
$$
ds^2=H_1^{-{1\over2}} H_2^{-{1\over2}}ds^2
(\bR^{(1,1)})+H_1^{{1\over2}} H_2^{-{1\over2}} ds^2(\bT^4)
+H_1^{{1\over2}}
H_2^{{1\over2}} ds^2(\bR^4)\ ,
\eqn\ddone
$$
where the directions 2345 are
compactified on $\bT^4$, and the harmonic functions, $H_1$ and $H_2$,
associated with the D-strings and D5-branes respectively, are given
by
$$
H_I=h_I+\sum_{A=1}^{N_I} {\lambda^I_{A}\over |{\vec x}-{\vec x}^{IA}|^2}\
.
\eqn\harmonic
$$
{}From the form of these functions, we see that the D-strings
have been organised into $N_1$ clusters, each consisting of $\lambda^1_A$,
$A=1,\cdots,N_1$ branes, with position in the transverse $\bR^4$ given by
$\vec{x}^{1A}$. Similarly, the D5-branes have been split into $N_2$ clusters,
consisting of $\lambda^2_A$, $A=1,\cdots,N_2$ branes with position
$\vec{x}^{2A}$. Clearly $Q_I=\sum_{A=1}^{N_I}\lambda^I_A$. Notice further that
the asymptotic volume $V$ of the torus $\bT^4$ as $|x|\rightarrow \infty$ is
given by,
$$
V=\sqrt{{\rm det} (h_1^{1/2} h_2^{-1/2}\,
\delta_{ab})}\rightarrow
{{h_1}\over {h_2}}\ ,
$$
where $a,b=2,\cdots,5$ label the coordinates of the
torus. Upon dimensional reduction on $\bT^4$ to six dimensions, this
supergravity solution becomes a string. We are interested in the moduli space
of such solitons, with the $4(N_1+N_2)$ positions $\vec{x}^{IA}$  considered
as collective coordinates.  The low-energy dynamics of these objects is then
described by a two-dimensional sigma model with (4,4) supersymmetry whose
target space is the moduli space. 
The computation of the moduli metric can be
done by adapting similar
results for black holes given in [\gutpapa]. We find,
$$
ds^2_{BH}= \sum_A( h_2 \lambda^1_{A} |d{\vec x}^{1A}|^2 +h_1 \lambda^2_{A}
|d{\vec x}^{2A}|^2)+\sum_{A,B} \lambda^1_{A}\lambda^2_{B} {|d{\vec
x}^{1A}-d{\vec x}^{2B}|^2 \over |{\vec x}^{1A}-{\vec x}^{2B}|^2}\ .
\eqn\modmetric
$$
This metric is compatible with two-dimensional 
(4,4)-supersymmetry if it is supplemented with an  appropriate torsion term
which in turn induces a Wess-Zumino term in the effective theory; the torsion
three-form is closed. The moduli space is a strong HKT manifold associated with
two hypercomplex structures  induced  from those on $\bR^4$ corresponding to a
basis of self-dual and anti-self-dual two-forms.
Note that the metric
\modmetric\ displays interaction terms only between branes of different type.
There are no two-derivative forces between branes of the same type, reflecting
the fact that in isolation each species of brane preserves 16 supersymmetries.
Such interactions would appear at fourth order in derivatives and it remains a
challenge to derive them through supergravity methods. Further note that in
the case that the D-strings are on top of the
D5-branes, $\vec{x}^{1A}=\vec{x}^{2A}$, and $\lambda^1_A=\lambda^2_A$,  the 
metric on the moduli space becomes that of Shiraishi [\shiraishi] (see
[\gutpapa] for $\lambda^1_A\not=\lambda^2_A$). 

\section{Gauge Theory} 

We
turn now to the gauge theory of the D1-D5 system. While attention is usually
focussed upon the Higgs branch of this theory, we will here be interested in
the Coulomb branch, parametrising the motion of the D1- and D5-branes in the
overall transverse space $\bR^4$. The gauge theory in question resides on the
$1+1$ dimensional intersection of the D1- and D5-branes, has ${\cal N}=(4,4)$
supersymmetry (eight supercharges) and gauge group $U(Q_1)\times U(Q_2)$ with
coupling constants $e_I$. The ratio of the coupling constants is determined by
the volume of the torus,
$$
e_1^2/e_2^2=V=h_1/h_2\ .
\eqn\coupling
$$
The matter
coupling consists of an adjoint hypermultiplet for each gauge group, together
with a single hypermultiplet in the bi-fundamental.
Let us focus on the bosonic
matter content of the above multiplets. For each of the vector multiplets, this
consists of a two-dimensional gauge field, together with four real adjoint
scalars which we will denote $\vec{\phi}^I$, where the index $I=1,2$ labels the
two gauge groups. Each hypermultiplet consists of a further four real scalars.
The vector multiplets and adjoint hypermultiplets arise from strings with both
ends on the D-string or both ends on the D5-branes, and furnish a
representation of ${\cal N}=(8,8)$ supersymmetry. This is reduced to ${\cal
N}=(4,4)$ by strings stretched between the D5 and D1-branes, giving rise to
the bi-fundamental hypermultiplet. 
While vacuum moduli spaces do not exist in
two dimensions, progress can still be made by deriving a low-energy sigma-model
description of the gauge theory in the spirit of a Born-Oppenheimer
approximation. The target space is then referred to as the vacuum moduli
space. Our theory has two branches of vacua: an $8(Q_1+Q_2)$ dimensional
Coulomb branch parametrised by the scalars in the two vector multiplets and
two adjoint hypermultiplets, and a $4Q_1Q_5$ dimensional Higgs branch
parametrised by the scalars in the adjoint and bi-fundamental hypermultiplets.
The latter is relevant when the D-strings are absorbed as instantons inside
the D5-brane. For the present purpose, it is the Coulomb branch that is of
interest. 
The vacuum expectation values of the scalars in the vector
multiplets parametrise the positions of the D-branes in 6789 directions, while
those of the scalars in the adjoint hypermultiplets determine the positions of
the D-strings and Wilson lines of the D5-branes in the 2345 directions. Upon
dimensional reduction to six dimensions, we will be interested only in the 6789
positions: the supergravity calculation does not capture the modes specified by
the adjoint hypermultiplet scalars. Thus we set the vacuum expectation values
of these scalars to zero and concentrate on the $4(Q_1+Q_2)$ dimensional
sub-manifold of the Coulomb branch parametrised by four adjoint scalars
$\vec{\phi}^{I}$.

The usual commutator terms in the scalar potential ensure
that $\vec{\phi}^I$ are  simultaneously diagonalisable. To compare to the
supergravity result, we further restrict attention to the $4(N_1+N_2)$
dimensional subspace of the Coulomb branch, on which  ${\vec \phi}^{I}={\rm
diag}\,({\vec \phi}^{IA})$, $A=1,\cdots,N_I$, where each entry ${\vec
\phi}^{IA}$ is proportional to the $(\lambda^I_A\times\lambda^I_A)$ unit
matrix. 
This results in the gauge symmetry breaking,
$U(Q_I)\rightarrow\prod_{A=1}^{N_I}U(\lambda^I_A)$. The existence of surviving
non-abelian gauge symmetries implies that this sub-manifold lies within a
singularity of the full Coulomb branch, reflecting the presence of these extra
massless excitations. Nonetheless, we may concentrate only on the subset of
deformations which preserve the form of the vacuum expectation value and derive
a low-energy effective action for these modes. 

Classically, this sub-manifold
of the  Coulomb branch is described by the flat metric
$({\bR}^{4N_1}/S_{N_1})\times ({\bR}^{4N_2}/S_{N_2})$, where the quotient
arises from the Weyl group of the gauge theory. The singularities correspond to
situations where the groups of D-strings or D5-branes  become coincident and
further non-abelian symmetry restoration occurs. 

The metric receives one-loop corrections from integrating
out massive matter, including W-bosons, off-diagonal terms of the adjoint
hypermultiplet, and the bi-fundamental hypermultiplets. Importantly, the
contribution from the first two of these cancel. This is obvious as together
they make a $(8,8)$-supersymmetric gauge multiplet and the moduli space metric
of any gauge theory with sixteen supercharges is constrained to be flat.  This
reflects the fact that the D1-branes and D5-branes do not interact at the
two-derivative level with branes of the same type. Thus the only corrections
come from the bi-fundamental hypermultiplets. Under the symmetry breaking, 
$U(Q_I)\rightarrow\prod_{A=1}^{N_I}U(\lambda^I_A)$, these decompose into
$N_1N_2$ hypermultiplets, each transforming in the bi-fundamental representation
of a single pair $U(\lambda^1_A)\times U(\lambda^2_B)$, $A=1,\cdots,N_1$ and
$B=1,\cdots,N_2$, and with mass $|{\vec \phi}^{1A}-{\vec \phi}^{2B}|$.  The
one-loop corrected Coulomb branch metric is given by
[\dls,\ds],
$$
ds^2_{gauge}= \sum_{A=1}^{N_1} {{\lambda^1_A}\over{e_1^2}}\
|d{\vec \phi}^{1A}|^2 + \sum_{B=1}^{N_2}{{\lambda^2_A}\over{e_2^2}}\ 
|d{\vec
\phi}^{2A}|^2  +\sum_{A=1}^{N_1}\sum_{B=1}^{N_2}\lambda^1_A\lambda^2_B
{{|d{\vec\phi}^{1A}-d{\vec\phi}^{2B}|^2}\over{|{\vec\phi}^{1A}-
{\vec\phi}^{2B}|^2}}\
,\eqn\dcoulomb
$$
where the factors of $\lambda$ in the first two, classical, terms 
come from tracing over block-diagonal matrices, and the third term arises from
integrating out the bi-fundamental hypermultiplets. In particular, the
$1/|mass|^2$ behaviour is typical of one-loop corrections in two-dimensional
gauge theories as may be seen by simple dimensional analysis. Notice that in
the simplest case in which the position of only a single D-string is allowed to
vary in the presence of fixed D5-branes, \dcoulomb\ reduces to the five-brane
metric [\dls,\ds]. The generalization of this result is that the full Coulomb
branch metric \dcoulomb\ coincides with the metric on the moduli space of the
D1-D5 system \modmetric\ if we identify $1/e_I^2=|\epsilon_{IJ}| h_J$.
This
further ensures that equation \coupling\ is satisfied.

As commented above,
supersymmetry requires that the metric be accompanied by a suitable torsion
term. Such terms are indeed generated at one-loop in the gauge theory [\ds] and
that the resulting low-energy dynamics is given by a two-dimensional
$(4,4)$-supersymmetric sigma-model.

We have shown that the classical moduli
space metric of five-dimensional black holes coincides with the one-loop
corrected Coulomb branch of an associated two-dimensional gauge theory. For
gauge groups of rank one, it can be argued that the restrictions of HKT,
together with $Spin(4)$ symmetry inherited from the R-symmetry of the gauge
theory, require that the metric receives no further corrections. While we know
of no such analysis for higher rank gauge groups, it seems plausible that
similar behaviour occurs. In particular, on the Coulomb branch in two
dimensions, there are no candidate instanton solutions to give semi-classical
non-perturbative corrections.

\chapter{The D2-D6 brane system}

\section{Supergravity}

The investigation of the D2-D6 brane
configuration is similar to that of the D1-D5 system of the previous section.
Indeed, the  two configurations are related by T-duality.
The D2-D6 brane
supergravity solution that we shall consider is
$$
ds^2=H_1^{-{1\over2}}
H_2^{-{1\over2}}ds^2 (\bR^{(1,2)})+
H_1^{{1\over2}} H_2^{-{1\over2}}
ds^2(\bT^4)+H_1^{{1\over2}} 
H_2^{{1\over2}} ds^2(\bR^3)\ ,
\eqn\daone
$$
where
$$
H_I=h_I+\sum_{A=1}^{N_I} {\lambda_{IA}\over |{\vec
x}-{\vec x}^{IA}|}\ ,
\eqn\datwo
$$
for $I=1,2$  are now harmonic functions on
$\bR^3$ associated with D2- and D6- branes, respectively. The moduli space
that
we shall examine is that parametrised by the positions  
$\vec{x}^{IA}$,
$I=1,2$, in the overall transverse  three-space of the D2- 
and D6-branes.

Upon reduction on $\bT^4$, we are left with a membrane type of solution
in six
dimensions and the effective theory is a three-dimensional sigma 
model with
eight supersymmetry charges. 
Supersymmetry requires that the sigma model
target
space is a HK manifold.
The moduli metric restricted on the positions
$\vec{x}^{IA}$, $I=1,2$, 
can be computed
by appropriately 
adapting the results
on black hole moduli spaces in [\gutpapa].
It was found that the moduli metric
is
$$
ds^2_{BH}= \sum_A( h_2 \lambda^1_{A} |d{\vec x}^{1A}|^2 
+h_1
\lambda^2_{A} |d{\vec x}^{2A}|^2)+\sum_{A,B} \lambda^1_{A}\lambda^2_B 
{|d{\vec
x}^{1A}-d{\vec x}^{2B}|^2 \over |{\vec x}^{1A}-{\vec x}^{2B}|}\ .
\eqn\again
$$
Note that, unlike for the 
D1-D5-brane configuration, we have not
identified all collective coordinates of the D2-D6 system. Indeed, the moduli
space  of
positions has dimension $3(N_1+N_2)$ while the moduli space of the
system
is expected to have $4(N_1+N_2)$ dimensions because it
must be HK. The
absence of these collective coordinates arises 
in the calculation of the moduli
metric because perturbations of 
high rank gauge potentials along the
worldvolume
of the D2-brane were ignored. Such perturbations
vanish in the black
hole case but they do not for the D2-D6 brane
configuration. However as we shall
review below, the moduli metric \again\ 
admits a unique hyperK\"ahler
completion by addition of $(N_1+N_2)$ 
periodic coordinates.

\section{Gauge Theory}

An analysis similar to that of the previous section may be given for
the gauge theory, which consists of a three-dimensional ${\cal N}=4$ (eight
supercharges) gauge multiplet with gauge group $U(Q_1)\times U(Q_2)$, a
hypermultiplet in the adjoint representation of the gauge group and a further
hypermultiplet in the bi-fundamental. While the bosonic matter content of the
hypermultiplet is unchanged in different dimensions, the vector multiplet now
contains a three dimensional gauge field and only three real, adjoint scalars
which we again denote as $\vec{\phi}^I$. Once again, when combined with the
adjoint hypermultiplets, these fields fill out a representation of the ${\cal
N}=8$ supersymmetry algebra and this is broken to ${\cal N}=4$ only by the
presence of the bi-fundamental hypermultiplet. Dimensional reduction of  this
theory to two-dimensions results in the model discussed in the previous
section.

Once again, the three dimensional gauge group is broken as 
$U(Q_I)\rightarrow \prod_{A=1}^{N_I} U(\lambda^I_A)$ and we restrict ourselves
to the relevant subspace of the Coulomb branch which is now of dimension
$3(N_1+N_2)$. The gauge theory supplies us with the remaining $(N_1+N_2)$
periodic scalars, $\sigma^I_A$, courtesy of the dual photons that are released
upon breaking the gauge group.

As in the two dimensional case, the one-loop
corrections from the adjoint hypermultiplets cancel those from the W-boson
multiplets, and only the bi-fundamental hypermultiplets contribute, making it
simple to immediately write down the one-loop corrected metric on the Coulomb
branch [\is], which is of the Lee-Weinberg-Yi type [\lwy] (see also [\gibb]),

$$
ds^2 = g_{AIBJ}d\vec{\phi}^{AI}\cdot
d\vec{\phi}^{BJ}
+(g^{-1})^{AIBJ}\psi_{AI}
\psi_{BJ}
\eqn\torichyp
$$
where the
first term is given by, 
$$
\sum_{A=1}^{N_1} {{\lambda^1_A}\over{e_1^2}}\
|d{\vec \phi}^{1A}|^2 + \sum_{B=1}^{N_2}{{\lambda^2_B}\over{e_2^2}}\ 
|d{\vec
\phi}^{2B}|^2  
+\sum_{A=1}^{N_1}\sum_{B=1}^{N_2}\lambda^1_A\lambda^2_B
{{|d{\vec\phi}^{1A}-d{\vec\phi}^{2B}|^2}\over{|{\vec\phi}^{1A}
-{\vec\phi}^{2B}|}}
\eqn\threed
$$
and
is seen to reproduce the supergravity result \again. Notice that, in contrast
to \dcoulomb, the one-loop correction in three dimensions has $1/|mass|$
behaviour. The second term in \torichyp\ is the hyperK\"ahler completion
mentioned above, with 
$$
\psi_{AI}=d\sigma_{AI} + \vec{\omega}_{AIBJ}
d\vec{\phi}^{BJ}\ ,
\eqn\toriii
$$
where $\vec{\omega}$ is defined by
$\nabla\times\vec{\omega}=\nabla g$. 
The manifold has $(N_1+N_2)$
tri-holomorphic isometries which act on the periodic coordinates $\sigma_{AI}$
by  constant shifts. Such a manifold is known as toric HK. These symmetries
are preserved within perturbation theory and the strong restriction of toric
hyperK\"ahlarity thereby ensures that there are no higher loop corrections to
the metric. However, instanton effects break this symmetry, and one may worry
about their presence. In three dimensions, the relevant semiclassical
configurations are monopoles. Importantly, in ${\cal N}=4$ three dimensional
gauge theories, and in contrast to their four-dimensional cousins, one-loop
effects around the background of the instanton do not cancel [\dkmtv].
Moreover, in gauge groups of rank $r\geq 2$, when the Coulomb branch is
interpreted in terms of soliton scattering these terms give rise to $r$-body
interactions [\cd]. As mentioned in the introduction, such interactions do not
appear from the supergravity perspective. To see that such terms do not appear
in the gauge theory either, one must determine whether instantons do indeed
contribute to the metric. In fact it is simpler to examine the four-fermi term
which is included in the supersymmetric completion of the metric. Instantons
can contribute to such a term only if they have precisely four fermionic zero
modes and no more. Thus, in order to determine whether instantons contribute in
the present case, we need only count fermionic zero modes. The necessary
observation is that the vector multiplet and adjoint hypermultiplet form an
${\cal N}=8$ multiplet which, for fundamental instantons, has $8$ zero modes;
too many to contribute to two derivative terms. One may wonder if four of these
can be lifted through couplings to the bi-fundamental hypermultiplets. However,
these hypermultiplets do not couple directly to the adjoint hypermultiplet, and
no such term can arise. Similar comments apply to the multi-instanton case.
Therefore instantons do not contribute to two derivative terms in these
theories, and the one-loop result \threed\ is exact.

We conclude this section
with the remark that the moduli metric of the D2-D6 brane system is T-dual to
the moduli metric of the D1-D5 brane system under Buscher type duality. This
can be seen by adapting the results of [\gary] to this case. It appears that
the Type II T-duality that relates the D2-D6 and D1-D5 brane systems induces
the Buscher T-duality on their moduli spaces.

\chapter{Probe Brane Potentials}

The action of a brane probe placed in a supergravity
background that preserves some supersymmetry is invariant, 
after gauge fixing   kappa-symmetry and worldvolume 
reparametrisations, under as many supersymmetry transformations as
those preserved by the background. However whether or not 
supersymmetry is preserved in certain phases of the theory depends
upon the existence of a supersymmetric ground configuration.
In many well-studied examples, including those of the previous 
sections, the derivative expansion of the probe 
action starts with velocity dependent terms. In such a situation, a  
supersymmetric configuration is achieved by simply ensuring that the probe 
is stationary and appropriately oriented with respect to the background. 
However there are many cases for which the derivative expansion starts with  
a potential term. In many cases
this potential arises due to the presence of a non-vanishing
expectation value for one or more fields of a p-brane action
compactified to a (p$-$n)-brane action.
The (p$-$n)-brane action then develops a potential with coefficient
that depends on the above expectation values. This is the
SS mechanism. The preservation of as many as  eight  supercharges in the probe 
action does not necessarily rule out such a potential [\agf, \gpaptown]. 
The issue then
is whether or not a supersymmetric configuration can be found which can
be interpreted as the supersymmetric  vacuum of the theory.

\section{SS Mechanism for transverse scalars}

We shall present two examples of brane action
compactifications with the SS-mechanism applied
to one of the transverse scalars. These involve
the M2- and M5-branes  in a KK-monopole background.
The former example has been already considered in [\BT]
but here we shall investigate the supersymmetric ground configuration
for the standard Taub/NUT metric.

For this we begin with the M-theory solution
of a KK-monopole  extended  in $0123456(10)$.
The corresponding supergravity solution is
$$
ds^2= ds^2(\bR^{(1,6)})+H^{-1}(d\theta+\omega)^2+ H ds^2(\bR^3)\ ,
\eqn\yesitiss
$$
where $H=1+{p\over |y|}$ is the harmonic function  on $\bR^3$, $dH={}^*d\omega$
and $y\in \bR^3$. The non-flat part of the metric is
the familiar Taub-NUT hyper-K\"ahler metric; the eleventh  coordinate
has been identified with $\theta=x^{10}$.
It is known that such solution preserves $1/2$ of the bulk
supersymmetry with supersymmetry projection
$$
\Gamma_7\Gamma_8\Gamma_9 \Gamma_{\theta}\epsilon=\epsilon\ .
\eqn\susycon
$$
In this background we place a
M2-brane probe and use static gauge in the
directions $012$. 
Next we compactify the direction $2$ on $S^1$ in such a way
that we keep only the zero modes for all field in all  directions
apart from that for the transverse scalar $\theta$
for which we set
$$
\partial_2\theta=q
\eqn\aatwoa
$$
where $q$ is a constant. This KK-ansatz is consistent
and it is a special case of that proposed in [\ss]. For compactifications
of the M2-brane see [\BT].
After integrating over $S^1$, the effective
action for such a system in the small derivative approximation is 
$$
\eqalign{
S= {1\over2}\int
 d^2x\ \big(&\delta_{ab}\eta^{\mu\nu} \partial_\mu z^a\partial_\nu z^b
+ H \delta_{ij} \eta^{\mu\nu} \partial_\mu y^i \partial_\nu y^j \cr 
&+ H^{-1}\eta^{\mu\nu} (\partial_\mu\theta+\omega_i \partial_\mu y^i)
(\partial_\nu\theta+\omega_j \partial_\nu y^j)
+q^2 H^{-1}\big)\ ,}
\eqn\actones
$$
where $\{z^a; a=1, \dots, 4\}$ are the transverse scalars in directions
$3456$,  $\{\theta, y^i; i=1,2,3\}$ are the three transverse
scalars associated with the KK-monopole (directions (10)789)
and $\mu,\nu=0,1$.
This action, apart from the standard kinetic term, 
also contains a potential which is the length of the
tri-holomorphic vector field of the Taub-NUT geometry. The lower dimensional
Lagrangian describes a string propagating is a KK-background
in the presence of a potential. As we shall see that there is
an alternative interpretation of the action \actones\
as describing a string propagating in the background of a
non-marginal ten-dimensional KK-monopole/ D6-brane background.

This system possesses a unique supersymmetric ground state in 
which a planar string lies at the origin of the KK-monopole, $y^i=0$, 
so the potential vanishes, and the rest of the transverse
scalars are constant. The supergravity solution associated
with the Taub/NUT metric
preserves $1/2$ of the bulk supersymmetry apart from near
the origin of the KK-monopole where all bulk supersymmetry
is preserved.   The string planar
worldvolume solution preserves $1/2$ of that of the background
and so $1/4$ of the bulk. 

Next let us examine
the theory near the origin of the KK-monopole.
The Taub/NUT metric near the origin is flat. In the
natural flat coordinates $(w^1, w^2, w^3, w^4)$, $|y|=|w|^2$.
In this case, the action \actones\ reduces to that
of a free theory with scalar potential
$$
V={q^2\over p} |w|^2\ ,
\eqn\abbtwob
$$
ie the fields along the KK-monopole directions are massive.
The supersymmetry preserved by the planar string solution
is $1/2$ of the bulk.

Apart from the string  solution above,  there 
 exists another supersymmetric solution to the classical 
equations of motion given by,
$$
\eqalign{
z^a&={\rm const}
\cr
y^i&={\rm const}\neq 0
\cr
\theta&=-q t\ ,}
\eqn\rotrot
$$
The solution \rotrot\  describes a string which rotates
with constant angular velocity $-q$ in the $\theta$ direction.
Observe that the solution is not invariant
under the string worldvolume Lorentz transformations.
The solution \rotrot\ can be easily lifted to a solution
for the M2-brane as follows:
$$
\eqalign{
z^a&={\rm const}
\cr
y^i&={\rm const}
\cr
\theta&=-q t+q x^2\ ,}
\eqn\rotrotm
$$
describing a M2-brane with the $x^2$ direction wrapped
on $\theta$ with winding number $q$, so $q\in \bZ$, and
rotating around $\theta$ with angular velocity $-q$.

To investigate the number of supersymmetry charges
preserved by the string  solution \rotrot, it is enough to
 investigate
the number of supersymmetry charges preserved by the lifted
M2-brane solution \rotrotm.
For this, we can use the supersymmetry condition associated
with the kappa-symmetry supersymmetry projection
[\becker, \kallosh, \stromk].
A brief calculation reveals that the supersymmetry projections
required are
$$
\eqalign{
\Gamma_0\Gamma_2\epsilon&=\epsilon
\cr
\Gamma_0\Gamma_1\Gamma_2\epsilon&=\epsilon\ .}
\eqn\ananana
$$
Therefore the solution \rotrot\ preserves $1/4$
of  supersymmetry of the  background or $1/8$
of the bulk as an immediate consequence of \susycon.

For our next example, we keep the same KK-monopole background as above,  but 
replace the M2-brane probe with  an M5-brane probe [\hsw] 
in the directions $012345$. 
We use static gauge and set the three-form self-dual
field of the M5-brane equal to zero because it does not
contribute to the potential in what follows.
Next we compactify the  M5-brane in the direction $x^5$ on $S^1$
and  keep only the zero modes for all transverse fields apart
from the transverse scalar $\theta$
for which we set
$$
\partial_5\theta=q
\eqn\eetaob
$$
where $q$ is a constant. In the small 
velocity approximation, the effective
action for such a system after integrating over $S^1$ is 
$$
\eqalign{
S= {1\over2}\int
 d^5x\ \big( &\eta^{\mu\nu} \partial_\mu z \partial_\nu z
+ H \delta_{ij} \eta^{\mu\nu} \partial_\mu y^i \partial_\nu y^j \cr 
&+ H^{-1}\eta^{\mu\nu} (\partial_\mu\theta+\omega_i \partial_\mu y^i)
(\partial_\nu\theta+\omega_j \partial_\nu y^j)
+q^2 H^{-1}\big)\ ,}
\eqn\actone
$$
where $z$ is the transverse scalar in the $x^6$ direction, $\{(\theta, y^i);
i=1,2,3\}$ are transverse scalars along the KK-monopole and 
$\mu,\nu=0,1,\dots,4$.
The action \actone\ describes a D4-brane in a KK-monopole
background which  apart from the standard kinetic term
for the transverse scalars also contains a potential as in
the string case above.

The analysis is now similar to the M2-brane probe. There exists a 
unique supersymmetric ground state in which the D4-brane lies 
at $y^i=0$ with all other transverse scalars constant.
 There further exists a 
classical solution of the D4-brane 
Lagrangian \actone\ given by,
$$
\eqalign{
z&={\rm const}
\cr
y^i&={\rm const}\neq 0
\cr
\theta&=-q t\ ,}
\eqn\rotrotd
$$
which can be lifted as a M5-brane solution as
$$
\eqalign{
z&={\rm const}
\cr
y^i&={\rm const}
\cr
\theta&=-q t+q x^5\ .}
\eqn\rotrotmd
$$
The solution \rotrotd\  describes a D4-brane which rotates
with constant angular velocity $-q$ in the $\theta$ direction
while the M5-brane wraps and rotates in the same direction.
Again both the D4 and M5-brane solutions are not
Lorentz invariant under the appropriate worldvolume
Lorentz transformations. The
supersymmetry projections associated with the M5-brane solution
above are
$$
\eqalign{
\Gamma_0\Gamma_5\epsilon&=\epsilon
\cr
\Gamma_0\Gamma_1\dots \Gamma_5\epsilon&=\epsilon\ .}
\eqn\eartwo
$$
Therefore using \susycon\ we find that 
the solution preserves $1/8$ of the bulk supersymmetry.

The above is clearly a special case 
of a more general class of constructions where
an M-brane is placed in a supergravity background with a
killing isometry. Then using T- and S-dualities, one
can construct D- and NS- brane actions with non-trivial
potentials. For example one can construct actions
with potentials  for all Dp-branes in
a KK-monopole background by compactifying or T-dualizing
the D4-brane action above.

It is also straightforward consider the case where the original background
involves many KK-monopoles  by allowing the harmonic function $H$ to have many
centres. In such a case the relevant action will also be given by \actone\
but now it will depend on the new harmonic function. In this case
apart of the solution that we have considered there are other Q-kink type
of solutions that preserve some supersymmetry; for work in this
direction see [\BT,\kinky ].

\section{Non-marginal BPS backgrounds and potentials}

There is an alternative interpretation for the
 action \actones\ as describing
a fundamental string propagating in a ten-dimensional
KK-monopole/D6-brane background. To illustrate this, we 
take the eleven-dimensional KK-monopole background
\yesitiss\  and change coordinates  as
$$
\eqalign{
\sigma&=q z+\theta
\cr
\rho&=z}
\eqn\transffff
$$
where $z$ is one of the coordinates in $\bR^{(1,6)}$ and $q$
is identified with the parameter in the SS mechanism for the
M2-brane.
Then we reduce the solution to ten-dimensions along $\rho$.
The resulting ten-dimensional solution is
$$
\eqalign{
ds^2&=(1+q^2H^{-1})^{{1\over2}} \big[ ds^2(\bR^{(1,5)}
+{1\over H+q^2} (d\sigma+\omega)^2+H ds^2(\bR^3)\big]
\cr 
e^{{4\over3}\phi}&=(1+q^2H^{-1})
\cr
A_1&= -H^{-1} (1+q^2H^{-1})^{-1} q (d\sigma+\omega)\ .}
\eqn\nonmarg
$$
In this background, we place a fundamental string and choose
a static gauge along a two dimensional subspace of $\bR^{(1,5)}$.
Now the effective Lagrangian of such a fundamental string in the
{\it small derivative} and {\it small} $q$ approximation coincides with that
of \actones\ after relabeling $\sigma=\theta$. In this 
approximation $q$ is in the same order as the derivatives
of the transverse scalars.

The above computation can be adapted easily for the case
of interpreting the result of the SS mechanism for the M5-brane.
In particular, the action \actone\ describes the dynamics
of a D4-brane probe in the background \nonmarg\
in the small derivative and small $q$ approximation.

In the case of D-branes a similar interpretation for the
actions with potential can be given. However in this case
the non-marginal background that it is probed is T-dual
to the one that we have started with. For later use, the
shall consider the SS reduction of the D3-brane in the
background of a ten-dimensional KK-monopole.
Changing coordinates as above and T-dualizing along the
direction $\rho$. The T-dual background is
$$
\eqalign{
ds^2&=ds^2(\bR^{(1,4)})+ (1+q^2 H^{-1})^{-1} d\rho^2
\cr &\qquad +{1\over H+q^2} (d\sigma+\omega)^2+H ds^2(\bR^3)
\cr 
e^{2\phi}&=(1+q^2H^{-1})^{-1}
\cr
H_3&= - d\big((1+q^2H^{-1})^{-1} d\rho \wedge (q d\sigma+\omega)\big)\ ,}
\eqn\nonmarg
$$
which describes a non-marginal ten-dimensional KK-monopole/NS5-brane
bound state. Probing this background with a D2-brane, 
the dynamics of the D2-brane is described in the small derivative and
small $q$ approximation by the action
$$
\eqalign{
S= {1\over2}\int
 d^3x\ \big( &\eta^{\mu\nu}\delta_{ab} \partial_\mu z^a \partial_\nu z^b
+\eta^{\mu\nu}\partial_\mu \rho \partial_\nu \rho
+ H \delta_{ij} \eta^{\mu\nu} \partial_\mu y^i \partial_\nu y^j \cr 
&+ H^{-1}\eta^{\mu\nu} (\partial_\mu\sigma+\omega_i \partial_\mu y^i)
(\partial_\nu\sigma +\omega_j \partial_\nu y^j)
+q^2 H^{-1}\big)\ ,}
\eqn\actone
$$
where $\{z^a; a,b=1,2\}$ are the scalars along $\bR^{(1,4)}$
transverse to the worldvolume directions of the D2-brane. Clearly
the FI parameter associated with the potential
 is determined by the expectation value
of a transverse scalar.

\section{SS  mechanism for Born-Infeld fields}

An alternative way to find
brane actions that exhibit a scalar potential is to
give an expectation value to a BI type of  field. 
This in particular can be applied in the case of
D-branes and for that of M5-branes. For the latter
case see also  [\keshav].

Here we shall investigate the case involving D6-brane probes 
in the D2-D6  brane system. The supergravity solution
for the D2-D6 brane system is
$$
\eqalign{
ds^2&= H^{-1} ds^2(\bR^{(1,2)})+ds^2(\bR^4)+H
ds^2(\bR^3)
\cr
e^\phi&=H^{-{1\over2}}
\cr
A_7&= dvol(\bR^{(1,2)}\oplus \bR^4) 
(H^{-1}-1)
\cr
A_3&= dvol(\bR^{(1,2)}) (H^{-1}-1)
}
\eqn\soltwo
$$
where
$\phi$ is the dilaton,
$A_3$
and $A_7$ are the R$\otimes$R gauge potentials associated with the D2-brane
and the D6-brane, respectively. We have also identify the
harmonic function of the D6-brane with that of the D2-brane. 
 For later use, the projections on the killing spinor associated
with the above background are
$$
\eqalign{
\Gamma_0\Gamma_1\dots \Gamma_6\epsilon&=-\epsilon
\cr
\Gamma_0\Gamma_1\Gamma_2\epsilon&=\epsilon\ .}
\eqn\susycont
$$

In this background we place a
D6-brane probe along the $\bR^{(1,2)}\oplus \bR^4$ directions
by choosing the static gauge. 
 The action of the probe is the standard Dirac-Born-Infeld (DBI)
one  including Chern-Simons terms.
In the small
derivative approximation  the DBI part of the action is
$$
\eqalign{
S_{BI}=\int
 d^7x\ \{-H^{-1}+1&+{1\over2} H \delta_{ij} \eta^{\mu\nu}
 \partial_\mu y^i \partial_\nu y^j+{1\over2}\delta_{ij}\delta^{ab}
 \partial_ay^i\partial_by^j
\cr &
+{1\over4}\big[ H^{-1} F_{\mu\nu} F^{\mu\nu}
+  2F_{\mu a} F^{\mu a}+
H^{-1}  F_{ab} F^{ab}\big]\}\ ,}
\eqn\actone
$$
where $y=y(x, z)$ are the three transverse
scalars, $F_{\mu\nu}$ and  $F_{ab}$ is the Born-Infeld (BI)
field in the directions 012 and  ($a,b=3456$),  respectively and
$F_{\mu a}$ are again the components of the BI field in the mixed
directions;
indices  are raised and lowered with respect to the flat metric.
The
contribution from the Chern-Simons term is
$$
S_{CS}=\int d^7x\ \{ H^{-1}-1+
{1\over4}(H^{-1}-1){ F}_{ab} {}^*{ F}^{ab}\}\ ,
\eqn\acttwo
$$
where the Hodge duality operation is with
respect to the
flat metric on $\bR^4$.
The two terms in the Chern-Simons
contributions come from
the D6-brane and D2-brane gauge potentials,
respectively.
Combining both terms we find
$$
\eqalign{
S={1\over2}\int d^7x\ \{& H \delta_{ij} 
\eta^{\mu\nu} \partial_\mu y^i \partial_\nu y^j+\delta_{ij}
\delta^{ab} \partial_a y^i
\partial_b y^j +
{1\over2} H^{-1} F_{\mu\nu} F^{\mu\nu}+ F_{\mu a} F^{\mu a}
\cr &
 -{1\over2}  F_{ab} {}^*F^{ab}
+{1\over2}H^{-1} [F_{ab} F^{ab}+ F_{ab}
{}^*F^{ab}]\}\ .}
\eqn\actthree
$$
The term involving the combination $(H^{-1}-1)$ cancels
between the BI and CS terms of the action because of the
BPS condition of the probe relative to the background [\atseytlin].

There are several ways to compactify the above action along the
directions $3456$ on $\bT^4$. For example one can perform
a standard $\bT^4$ torus compactification. The resulting action
will be that of a D2-brane propagating in the background\foot{
Strictly speaking the compactification should be followed
by a T-duality on the background. However \soltwo\ is invariant
under T-duality along all directions on $\bT^4$. }
Alternatively, one can perform a non-trivial compactification
by allowing 
$$
F_{ab}=B_{ab}
\eqn\ttwtwooo
$$
where $B$ is a constant field.
In such case the resulting action is
$$
\eqalign{
S={1\over2}\int d^3x\ \{& H \delta_{ij} 
\eta^{\mu\nu} \partial_\mu y^i \partial_\nu y^j+\delta_{ij}
\delta^{ab} \partial_a y^i
\partial_b y^j +
{1\over2} H^{-1} F_{\mu\nu} F^{\mu\nu}+
\delta_{ab}\eta^{\mu\nu} \partial_\mu z^a \partial_\nu z^b
\cr &
 -{1\over2}  B_{ab} {}^*B^{ab}
+{1\over2}H^{-1} [B_{ab} B^{ab}+ B_{ab}
{}^*B^{ab}]\}\ ,}
\eqn\ractthree
$$
where $\{z^a\}$ are the Kaluza-Klein scalars associated with the
BI field.
This action again described a D2-brane in the background \soltwo\ but also
exhibits a scalar potential with coefficient dependent
on the non-vanishing expectation value of the BI field. The potential
term vanishes if $B$ is chosen to be anti-self-dual. The above
compactification followed by an appropriate truncation is
consistent, ie solutions of the reduced action are also
solutions of the higher dimensional one.

Let us now discuss the supersymmetric configurations of this probe 
brane action. Firstly let us suppose that $B=B^-$ is anti-self-dual.
In such a case the potential vanishes. A solution of the system 
is that of standard planar D2-brane located at a point $y^i={\rm const}$
and $z^a={\rm const}$ in the background with $F_{\mu\nu}=0$. Such solution
preserves $1/4$ of the supersymmetry. A non-vanishing value
of $B$ does affect the number of supersymmetries
preserved by the configuration. This can be easily seen by
lifting this solution to that of a planar D6-brane probe
and then use the supersymmetry projector arising from kappa-symmetry
 [\kallosh, \stromk].
The supersymmetry projector associated with $B=B^-$ is the same
as that of the D2-brane of the background. One may view the effect 
of $B$ as inducing more D2-brane
charge on the original D6-brane. These new D2-branes
lie parallel to those of the background and so no more supersymmetry
is broken.

For another supersymmetric configuration, we decompose
 $B=B^++B^-$ into 
self-dual and anti-self-dual parts, we require that $B^+\neq 0$. 
Moreover we seek a solution for which  $z^a={\rm const}$ and
$A_\mu=A_\mu(x)$ and $y^i=y^i(x)$.
Substituting these into the remaining field equations,
we find
$$
\eqalign{
\partial_\mu( H\partial^\mu y^i)-
{1\over2}\partial^i H^{-1} (B^+_{ab} B^+{}^{ab})&=0
\cr 
\partial_\mu (H^{-1} F^{\mu\nu})&=0\ .}
\eqn\yer
$$
A solution for this system is
$$
\eqalign{
y^i&=0
\cr
F_{\mu\nu}&={\rm const}\ .}
\eqn\yert
$$
This is the most general vacuum configuration in this sector.
The investigation of supersymmetry in more subtle. The
background we are considering \soltwo\ does not have
a well defined near horizon geometry as $|y|\rightarrow 0$.
Consequently, the killing spinors are not well defined at that
point. However  since for a generic point the background
preserves $1/4$ of supersymmetry, the effective theory
preserves eight supersymmetry charges by
continuity one may  argue that the same number of supersymmetry charges
survives at $y=0$. Assuming this, we take 
$B^+_{34}=B^+_{56}\not=0$, $F_{12}\not=0$ non-zero and with the
rest of the components to vanish. Lifting the D2-brane solution
to that of the D6-brane probe, the naive supersymmetry conditions
arising  from kappa-symmetry become after 
using the projections \susycon\ the following:
$$
{1\over2}{\cal F}_{MN} \Gamma^M\Gamma^N\epsilon= [F_{12} \Gamma^1\Gamma^2+
B^+_{34}(\Gamma^3\Gamma^4+\Gamma^5\Gamma^6)]\epsilon=0\ .
\eqn\actfive
$$
This  can be rewritten using again \susycon\ as
$$
F_{12} \Gamma^1\Gamma^2+2B^+_{34}\Gamma^3\Gamma^4=0\ ;
\eqn\actsix
$$
Observe that the $B^-$ part does not contribute in the
supersymmetry condition above.
This leads to a supersymmetry projection provided
$$
F_{12}=\pm 2B^+_{34}\ .
\eqn\actseven
$$
Therefore the configuration preserves $1/2$ of that of the background, 
and $1/8$ of the bulk.  
Let us discuss the possible interpretation of this bound state 
in the bulk. 
This configuration clearly involves D2 and D6 branes. However, one 
may also view $B^+$ as inducing anti-D2-brane charges arising from $B^+$. 
Finally $F_{12}$ induces,
using standard arguments (see e.g. [\douglas, \pol])
 after further compactifying
on $\bT^2$, D0-brane charges on the probe.
Therefore the bulk configuration should have the  
interpretation of a D0-D2-${\bar {\rm D2}}$-D6 bound state. However, the 
existence of a BPS solution of the effective theory does not necessarily 
imply the existence of a bound state in the full string theory, as 
one also expects tachyonic modes to be present in the system which 
have not been taken into account in the above analysis [\sw].

A similar analysis can be done for the
D1- D5-brane system leading to similar 
conclusion but now involving
a bound state of a D-instanton, a D-string, 
a anti-D-string and a D5-brane.
The relevant action in this case is as 
in \ractthree\ but there are some differences.
One difference is  that the harmonic function $H$ which  appears in the
action  is that 
on $\bR^4$ instead of $\bR^3$  and another is that the integration
in the same action is over
a string worldvolume.
In addition  the D1-D5 background considered
here  has near horizon geometry
$AdS_3\times S^3\times \bR^4$ preserving
$1/2$ of the bulk supersymmetry [\sked].
 Note also that
in order to introduce D-instantons one has to consider
the Euclidean DBI action.

It is also straightforward consider the case 
where the original background
involves many D2-D6 branes by allowing the 
harmonic function $H$ to have many
centres. In such a case the relevant 
action will also be given by \actthree\
but now the harmonic function will have many centres. In this case
apart of the solution that we have 
considered there are other Q-kink type
of solutions that preserve 
some supersymmetry. It would be of interest to
investigate  these solution further; see also [\BT,\kinky ].
The above analysis can also be carried out with the
full non-linear DBI action.

\section{Potentials in Gauge Theory}

In this final section, we discuss the generation of potentials 
on the moduli spaces of gauge theories. Specifically, we will 
return to the D2-D6 system of section 2. It will suffice 
to consider a single D2 and $N$ D6-branes. We take the volume of the
torus $\bT^4$ to be infinite which ensures 
$e_2\rightarrow\infty$ and the dynamics of the D6-branes are frozen. 
The moduli space metric \torichyp\ reduces 
to the Coulomb branch of the D2-brane worldvolume theory 
which is given by the multi-centred Taub-NUT space, 
$$
ds^2 = Hd\vec{\phi}\cdot
d\vec{\phi}+H^{-1}
({\rm d}\sigma+\vec{\omega}\cdot{\rm d}\vec{\phi})^2
\eqn\target
$$
with
$H=1/e_1^2+N/|\vec{\phi}|$, where for $N>1$, the metric is singular. 
This reflects the fact that the D2-brane is a probe 
in the multi-Kaluza-Klein monopole geometry of the 
D6-branes. In particular, the dual photon $\sigma$ may be identified 
with the eleventh dimension. For $N>1$, the gauge theory has a 
Higgs branch emanating from the origin of the Coulomb branch. 

Now let us
consider how things change when we introduce a Fayet-Iliopoulos 
(FI) parameter, $\zeta$. Classically these terms ensure that the 
Coulomb branch of the gauge theory no longer exists. However, one may nevertheless 
derive a description of the low-energy dynamics of the vector multiplet as 
a massive sigma model with target space \target and a potential energy $U$. 
From the classical lagrangian, the potential on the Coulomb branch is 
given by $\ft12 e_1^2\zeta^2$. However, in the full theory 
the coupling constant $e_1$ is replaced by its quantum corrected 
value, resulting in
$$
U=\ft12  H^{-1}\zeta^2\ .
\eqn\pot
$$

This potential is familiar from the preceding sections; it has 
has a minimum at $\phi=0$, implying an induced 
attractive force between the D2 and D6-brane. 
Note that the resulting  dynamics preserves eight supercharges as can 
be seen by noting that the potential is
proportional to the length of 
a tri-holomorphic Killing vector associated with
the isometry $\sigma\rightarrow\sigma+c$ [\agf, \gpaptown]. 

It has been argued that the bulk interpretation of the FI parameter 
is as a self-dual background NS$\otimes$NS B-field [\abs,\sw]. 
Indeed, open string calculations reveal an attractive force 
between the D2 and D6-branes in the presence of a B-field.  
The main evidence for the specific identification of the self-dual 
part of the B-field with the FI parameter comes from looking not at the 
Coulomb branch as above, but at the Higgs branch. For theories with 
several D6-branes, the Higgs branch is deformed by the FI parameter 
into the moduli space of a single non-commutative instanton in 
$U(N)$ gauge theory. This agrees with the string theory picture 
of the D2-brane dissolving as an instanton in the D6-branes which, 
in the presence of a background B-field, support a non-commutative 
Yang-Mills theory. 

Although the field theory has a unique supersymmetric vacua, 
as in the previous sections, there are further classical 
supersymmetric solutions, given by, 
$$
\vec{\phi}=\vec{\phi}_0={\rm const.}\neq 0 \
\ \ \ ,\ \ \ \ \dot{\sigma}= \zeta
\eqn\sol
$$
which may be
simply seen to be a BPS solution by completing the 
square in the Hamiltonian
and noting that the residual cross-term 
is the Noether charge, $Q$, associated
with the isometry that shifts 
$\sigma$. The energy of this state is thus
$E=H^{-1}\zeta^2=Q$, 
and is seen to be correlated with the 
separation of the D2-brane from the D6-brane. 

As mentioned previously, the interpretation of these states 
in the full IIA string theory is unclear due to issues 
associated with tachyons. However, in this case, there does indeed 
exist a natural interpretation. One may dualise $\sigma$ into the three 
dimensional field strength $F$ which, for the above  solution, gives, 
$$
F_{12} = H^{-1} \dot{\sigma}=\zeta H^{-1}
\eqn\fonetwoa
$$
Alternatively, one may consider this to be a non-marginal 
D0-D2 bound state. This state therefore has the interpretation of a D0-D2-D6 
bound 
state in the background of a constant NS$\otimes$NS B-field. It 
preserves $1/8$ of the supersymmetry of the bulk.

\vskip 1cm
\noindent{\bf
Acknowledgments:}  
We would like to thank M. Aganagic, S. Das, J. Gauntlett, A. Hanany,
A. Karch, J. Minahan and W. Taylor for useful discussions. G.P. would 
like to thank CERN
for a research associateship and CALTECH for hospitality while this work was in
progress. D.T. would like to thank the Center for Theoretical Physics, MIT, 
Boston and S. Mukhi and the Department of Theoretical Physics, TIFR, Mumbai 
for hospitality while this work was completed. G.P. is supported by a 
University Research Fellowship from the
Royal Society. D.T. is supported by an EPSRC fellowship. Both authors
acknowledge  SPG grant PPA/G/S/1998/00613 for further support. This work is
supported in part by funds provided by the U.S. Department of Energy (D.O.E)
under cooperative research agreement \# DF-FCO2-94ER40818. 

\refout
\end